\documentclass[fleqn,twoside]{article}
\usepackage{espcrc2}
\usepackage{amssymb}

\usepackage{graphicx}

\title{HEAVY QUARKONIA -- A Review of the Experimental Status}

\author{Kamal K. Seth\address[MCSD]{Department of Physics and Astronomy, 
        Northwestern University\\Evanston, IL, 60208, USA}}
       
\begin{document}

\begin{abstract}
A review of the present status of the spectroscopy of heavy quarkonia ($b\bar{b}$, $c\bar{c}$) is presented.
\end{abstract}

\maketitle

\section{Introduction}

Because strong interactions are flavour-independent, it would appear that they can be equally well studied in the spectroscopy of light or heavy quarks.  However, the highly relativistic nature of quarks in light-quark ($u,d,s$) hadrons, the large value of the strong coupling constant, $\alpha_s$ ($\gtrsim0.6$), and the near equality of the masses of the $u,d,s$ quarks, makes the light quark states strongly overlapping, with small spacing $\approx14$ MeV, large widths, $\Gamma\approx150$ MeV, and mostly mixtures of all three flavours.  In contrast, $b\bar{b}$ and $c\bar{c}$ hadrons are relatively free from these problems, with $\left<v^2/c^2\right>\approx0.1-0.2$, $\alpha_s=0.2-0.3$, and have well resolved states (spacing$\approx15-40$ MeV, $\Gamma\approx0.05-5$ MeV).  The spectra of $|c\bar{c}>$ charmonium and $|b\bar{b}>$ bottominium are illustrated in Fig. 1.

The history of quarkonia began with the discovery of $J/\psi$ in 1974 and lots of $c\bar{c}$ charmonium discovery physics was done by SLAC, DESY, and Orsay during the next ten years.  Because only vector states can be directly formed in $e^+e^-$ annihilations, in these experiments precision results could be obtained for vector states ($J/\psi$, $\psi'$), but this was often not the case with other states ($^{1,3}P_J$, $^1S_0$).  This shortcoming was at least partially removed by charmonium spectroscopy with $p\bar{p}$ annihilation in which states of all $J^{PC}$ could be  directly formed.  In the Fermilab experiments E760/E835 precision mass and width measurements were made of $^3P_J$ states ($\chi_{c0}$, $\chi_{c1}$, $\chi_{c2}$).  Unfortunately, the limitations imposed by the available luminosity, and the absence of charged particle tracking limited even these experiments in the pursuit of the spectroscopy of spin-singlet states, $\eta_c(1^1S_0)$, $\eta_c'(2^1S_0)$, $h_c(1^1P_1)$.  In 1989 the BEPC (Beijing) brought large $e^+e^-$ luminosity to bear on the spectroscopy of charmonium, and its BES detector has made notable contributations to improving the precision achieved in earlier $e^+e^-$ experiments.  The spin-singlet states still remained mostly out of the reach of BES.  A very favourable development has, however, taken place in the last few years.  The CESR accelerator at Cornell, which had operated since 1979 in the 9--11 GeV region, and had made important contributions in the spectroscopy of $b\bar{b}$ bottomonium, has been converted to CESR-c, designed to work optimally in the 3--5 GeV region. It is beginning to make measurements in the charmonium region with the great advantage offered by its excellent detector which has been upgraded from CLEO II to III to CLEO-c.  Also the huge $e^+e^-$ luminosities  available at KEK (Belle detector) and SLAC (BaBar detector) are making it possible to produce very competative results in charmominium spectroscopy, even as they run exclusively at $\Upsilon(4S)$.  To complete this historical narrative, let me mention that at GSI (Darmstadt) a dedicated facility for $p\bar{p}$ experiments in the $\sqrt{s}=3-5$ GeV region is being built, and in Beijing the construction of BEPC-II and BES-III has been approved.  We can therefore look forward to a very bright future for the spectroscopy of this mass region, which not only includes charmonium but also the QCD exotics, glueballs, hybrids, and whatever else lies there hidden.

\begin{figure*}[bt]
\includegraphics[width=6.in]{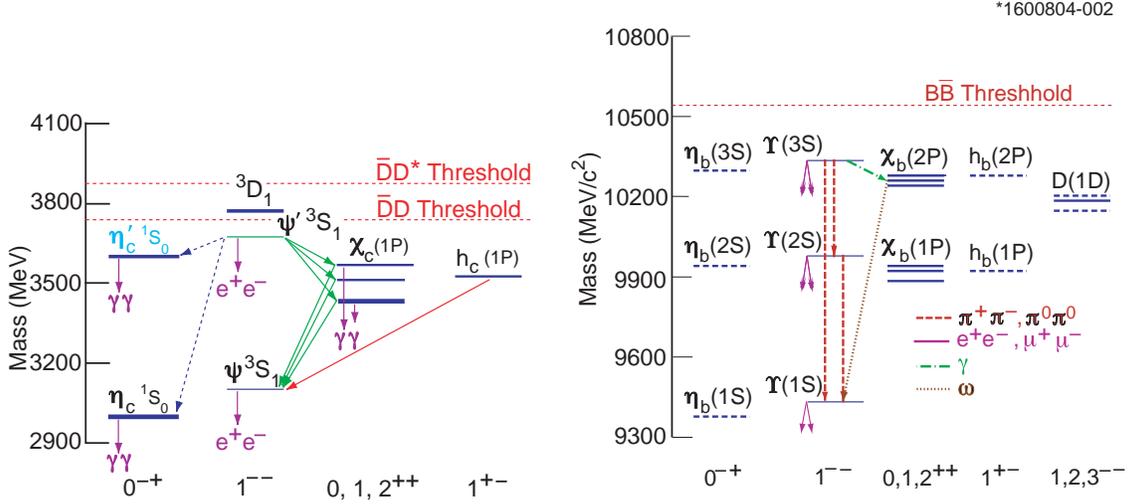}
\caption{Spectra of quarkonium states, (left) charmonium, (right) bottomonium.}
\end{figure*}

\section{Bottomonium}

Despite the fact that the bottomonium $b\bar{b}$ system is certainly more amenable to pQCD, we know far less about bottomonium than we know about charmonium.  The $\eta_b$, ground state of bottomonium, has not been identified so far.  The vector states $\Upsilon(1S, 2S, 3S$ and $4S)$ are known but only one hadronic transition from these, $\Upsilon(nS)\to\Upsilon(n'S)\pi^+\pi^-,\; (n'<n)$ has ever been observed.  Radiative transitions $\Upsilon(nS)\to\gamma\chi_b(n'^3P)$ states have been observed.  No D-states, which are expected to be bound (see Fig. 1) have been observed.  No hadronic transition from any $\chi_b$ states has ever been observed.  Recently, CLEO has made small gains in both the above problems. The $1^3D_2$ state has been successfully observed in 4-photon cascade $\Upsilon(3S)\to\gamma_1(2P)\to\gamma_1\gamma_2(1D)\to\gamma_1\gamma_2\gamma_3(1P)\to\gamma_1\gamma_2\gamma_3\gamma_4\Upsilon(1S),\Upsilon(1S)\to l^+l^-$.  The mass $M(1^3D_2)=10,161.1\pm0.6\pm1.6$ MeV\cite{cleob}.  In another measurement, $\Upsilon(3S)\to\gamma\chi_b(2P),\chi_b(1,2)\to\omega\Upsilon(1S)$ has also been observed\cite{cleoc}.

In addition, precision measurements have been made by CLEO \cite{cleog} for the leptonic branching fractions of $\Upsilon(1S,2S,3S)$ states.  It is found that while $\mathcal{B}(\Upsilon(1S)\to\mu^+\mu^-)$ (see Table 1) is in agreement with the current PDG \cite{pdg} average,  those for $\Upsilon(2S)$ and $\Upsilon(3S)$ are 55\% and 35\% larger, respectively.  This leads to a corresponding decrease in total widths.  Precision measurements of radiative decays of $\Upsilon(2S)\to\gamma\chi_b(1P)$, $\Upsilon(3S)\to\gamma\chi_b(2P)$ are also being made.

\begin{table*}
\begin{center}
\begin{tabular}{|r|c|c|c|c|}
\hline
 & $B_{\mu\mu}(\%)$ CLEO & $B_{\mu\mu}(\%)$ PDG & $\Gamma$ (keV) CLEO &  $\Gamma$ (keV) PDG \\
\hline
$\Upsilon(1S)$ & $2.49\pm0.02\pm0.07$ & $2.48\pm0.06$ & $52.8\pm1.8$ & $53.0\pm1.5$ \\
$\Upsilon(2S)$ & $2.03\pm0.03\pm0.08$ & $1.31\pm0.21$ & $29.0\pm1.6$ & $43\pm6$ \\
$\Upsilon(3S)$ & $2.39\pm0.07\pm0.10$ & $1.81\pm0.17$ & $20.3\pm2.1$ & $26.3\pm3.4$ \\
\hline
\end{tabular}
\end{center}
\caption{New CLEO results for muonic branching ratios and total widths of $\Upsilon(1S,2S,3S)$ resonances.}
\end{table*}

\section{Charmonium}

\subsection{The $\rho-\pi$ Problem}

Since the widths for leptonic decays, as well as 3 gluon decays to light hadrons, of both $J/\psi$ and $\psi'$ depend on the wave functions at the origin, pQCD predicts the equality of the ratios of branching ratios
$$\frac{B(\psi'\to l^+l^-)}{B(J/\psi\to l^+l^-)} = (13\pm2)\% = \frac{B(\psi'\to LH)}{B(J/\psi\to LH)}.$$
This expectation has been extended to ratios of individual hadronic decays, and has led to many measurements by BES and CLEO \cite{cleox} to test it.  The results is that while the sums of all hadronic decays do seem to follow this expectation, and $\sum_i B(\psi'\to LH)_i/\sum_i B(J/\psi\to LH)_i = (17\pm3)\%$, individual decays show large departures from it, the ratio being as small as 0.2\% for $\rho\pi$ decays.  While many exotic theoretical suggestions have been made to explain these deviations, it appears that what we are witnessing is the failure of attempts to stretch pQCD beyond its limits of validity.

\subsection{The Spin Singlet States and the Hyperfine Interaction}

The spin-indepedent $q\bar{q}$ interaction is well understood in terms of one-gluon exchange, and is very successfully modeled by a Coulombic 1/$r$ potential.  The spin dependence which follows from this is also accepted.  What is not understood is the the nature of the confinement part of the interaction, which is generally modeled by a scalar potential proportional to $r$.  A crucial test of the Lorenz nature of the confinement potential is provided by the measurement of hyperfine or spin-singlet/spin-triplet splittings.  A scalar potential does not contribute to the spin-spin or hyperfine interaction, whereas for a Coulombic potential it is a contact interaction.  As a consequence hyperfine splitting is predicted to be finite only for S-wave states, and to be zero for P-wave and higher L-states

\subsubsection{Hyperfine Splitting in S-wave Quarkonia}

No singlet states have so far been identified in bottomonium.  In charmonium, however, it has been established for a long time that $\Delta M(1S)_{hf}\equiv M(J/\psi, 1^3S_1)-M(\eta_c, 1^1S_0)=172\pm2$ MeV.  It is interesting to determine the size of the hyperfine splitting of $2S$ states, which sample the confinement region more deeply.  Long ago Crystal Ball\cite{cball} claimed the identification of $\eta_c'$ with $M(\eta_c')=3594\pm5$ MeV, leading to $\Delta M(2S)_{hf}=92\pm5$ MeV, which kind of made sense with $\Delta M(1S)_{hf}=172\pm2$ MeV.  Most potential model calculations tried to accomodate this `experimental' result , although it was not confirmed by any subsequent measurement, and was actually dropped by the PDG meson summary.

The seach for $\eta_c'$ has finally ended. Belle\cite{bellea} announced it first in two different decays of large samples of B-mesons.  CLEO\cite{cleoa} and BaBar\cite{babara} both have identified it in the two-photon fusion reaction, $e^+e^-\to(e^+e^-)\gamma\gamma,\gamma\gamma\to\eta_c'\to K_SK\pi$.  The CLEO measurement is shown in Fig. 2. The exciting part of these measurements is that $M(\eta_c')_{avg}=3637.4\pm4.4$ MeV, which is almost 50 MeV larger than the old Crystal Ball claim, and it leads to a surprisingly small hyperfine splitting, $\Delta M(2S)_{hf}=48.6\pm4.4$ MeV.  It is too early to say whether this can be explained in terms of channel mixing\cite{elq}, or unexpected contribution from the confinement potential.

\begin{figure}[tb]
\centerline{\includegraphics[width=3.in]{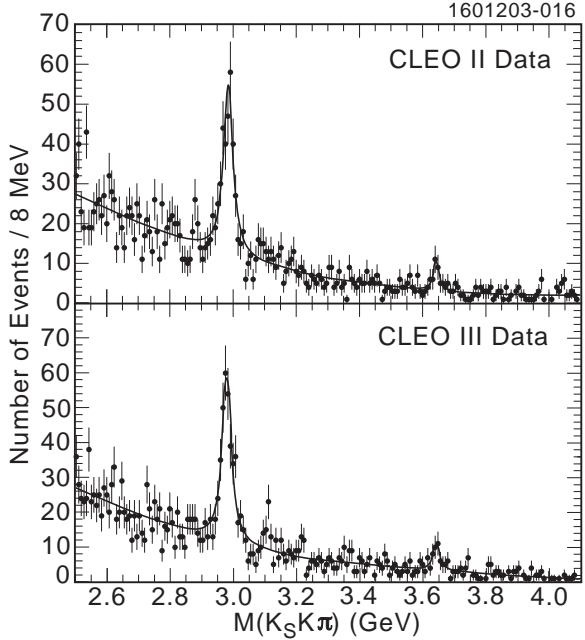}}
\caption{CLEO discovery of $\eta_c'(2^1S_0)$ in two photon formation, and decay into $K_S K^\pm\pi^\mp$.}
\end{figure}

\subsubsection{Hyperfine Splitting in P-wave Quarkonia}

As mentioned already, hyperfine splitting is expected to be zero in all except S-wave states if the confinement potential is scalar, as is generally assumed.  Thus it is expected that $\Delta M(1P)_{hf} \equiv \left<M(1^3P_J)\right>-M(1^1P_1)=0$, except for higher order contributions of no more than an MeV or two.  Unfortunately, while $\left<M(1^3P_J)\right>=3525.31\pm0.07$\cite{cester}, the $h_c(1^1P_1)$ has not been firmly identified.  Let me however, give you a preview of the present situation.  Both Fermilab E835 and CLEO are working on the search for $h_c$.  The E835 experiment is analyzing the reactions $p\bar{p}\to h_c\to\pi^0 J/\psi$ and $p\bar{p}\to h_c\to\gamma\eta_c$, and preliminary results are that while the first reaction does not have a signal for $h_c$ formation\cite{dave}, the second may have.  The CLEO team is analyzing $e^+e^-\to\psi'\to\pi^0 h_c,h_c\to\gamma\eta_c$ but has not presented any results so far [Note: Since the conference, CLEO has announced its preliminary results with $M(h_c)=3524.8\pm0.7$ MeV\cite{amiran} with the consequent $\Delta M(1P)_{hf}=0.6\pm0.6$ MeV. It appears that there is no significant departure from the simple expectation, $\Delta M(1P)_{hf}$=0].

\subsubsection{Higher Vector States}

For a long time the parameters listed in the PDG compilation for the three vector states above the $D\bar{D}$ threshhold have been based on the R-parameter measurement by the DASP group\cite{dasp}, even though none of the other measurements of R agreed with it.  Recent measurements by the BES group\cite{besa} have finally allowed us\cite{sethb} to make a reliable determination of the masses, total widths, and leptonic widths of these states.  Fig. 3 shows fits to the BES \cite{besa} and CB \cite{cball} data, and Table 2 lists the results for the parameters.

\begin{table}
\begin{small}
\begin{tabular}{|c|c|c|c|}
\hline
\hline
 & $M^{(1)}$ & $\Gamma^{(1)}_{tot}$ & $\Gamma^{(1)}_{ee}$ \\
 & (MeV) & (MeV) & (keV) \\
\hline
PDG\cite{pdg} & $4040 \pm 10$ & $52 \pm 10$ & $0.75 \pm 0.15$ \\
CB\cite{cball} & $4037 \pm 2$ & $85 \pm 10$ & $0.88 \pm 0.11$ \\
BES\cite{besa} & $4040 \pm 1$ & $89 \pm 6$ & $0.91 \pm 0.13$ \\
\hline
CB+BES & $4039.4 \pm 0.9$ & $88 \pm 5$ & $0.89 \pm 0.08$ \\
\hline
\hline
 & $M^{(2)}$ & $\Gamma^{(2)}_{tot}$ & $\Gamma^{(2)}_{ee}$ \\
\hline
PDG\cite{pdg} & $4159 \pm 20$ & $78 \pm 20$ & $0.77 \pm 0.23$ \\
CB\cite{cball} & $4151 \pm 4$ & $107 \pm 10$ & $0.83 \pm 0.08$ \\
BES\cite{besa} & $4155 \pm 5$ & $107 \pm 16$ & $0.84 \pm 0.13$ \\
\hline
CB+BES & $4153 \pm 3$ & $107 \pm 8$ &  $0.83 \pm 0.07$\\
\hline
\hline
 & $M^{(3)}$ & $\Gamma^{(3)}_{tot}$ & $\Gamma^{(3)}_{ee}$ \\
\hline
PDG\cite{pdg} & $4415 \pm 6$ & $43 \pm 15$ & $0.47 \pm 0.10$ \\
CB\cite{cball} & $4425 \pm 6$ & $119 \pm 16$ & $0.72 \pm 0.11$ \\
BES\cite{besa} & $4429 \pm 9$ & $118 \pm 35$ & $0.64 \pm 0.23$ \\
\hline
CB+BES & $4426 \pm 5$ & $119 \pm 15$ &  $0.71 \pm 0.10$\\
\hline
\hline
\end{tabular}
\end{small}
\caption{Summary of results from ref. \cite{sethb}. Masses $M^{(i)}$ and total widths $\Gamma_{tot}^{(i)}$ are in MeV, electron widths $\Gamma_{ee}^{(i)}$ are in keV.}
\end{table}

\begin{figure}[tb]
\centerline{\includegraphics[width=3.5in]{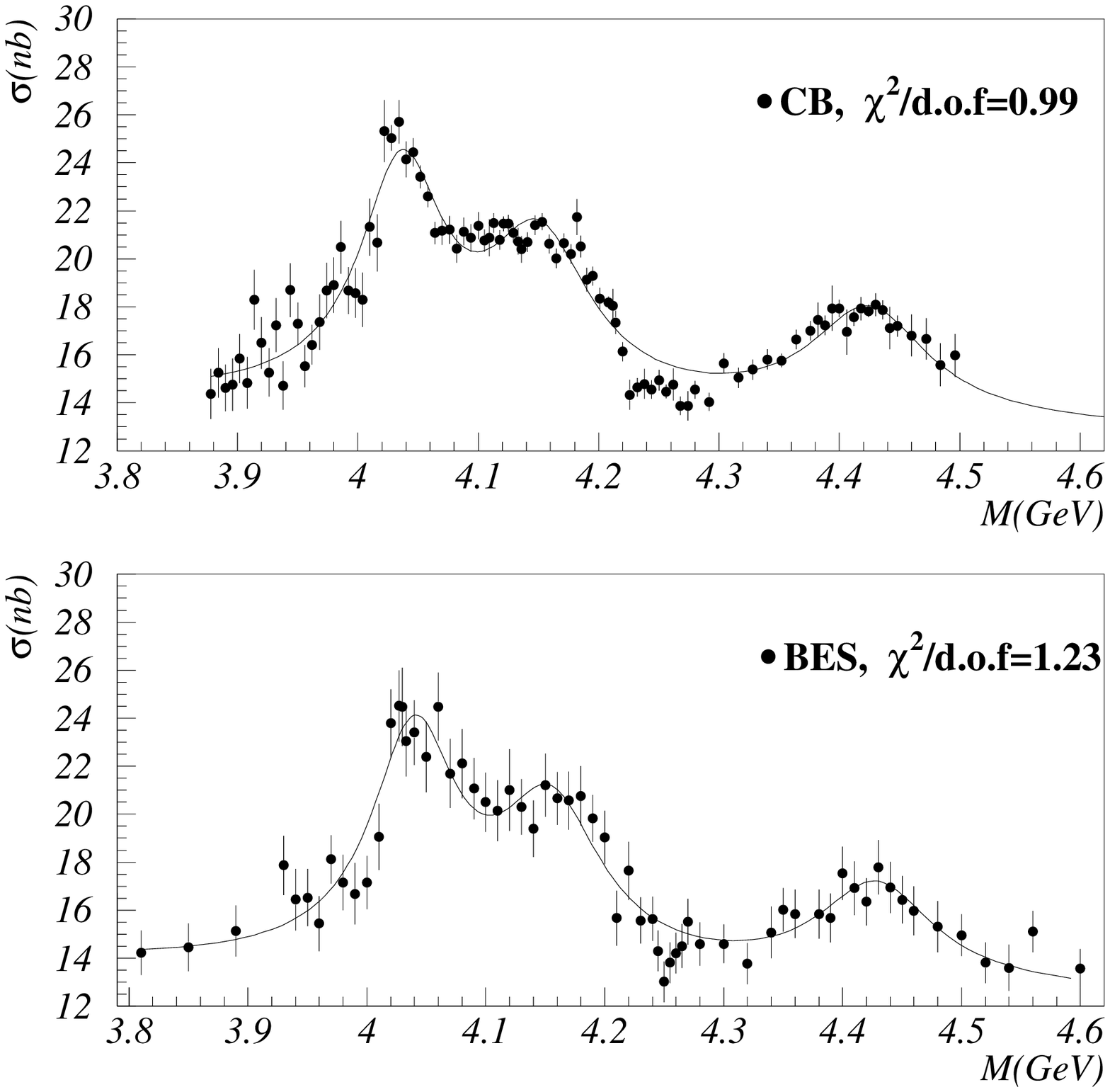}}
\caption{$R\equiv\sigma(e^+e^-\to\mathrm{hadrons})/\sigma(e^+e^-\to\mathrm{leptons})$ from BES \cite{besa} and Crystal Ball \cite{cball}.  The fits are from ref. \cite{sethb}.}
\end{figure}

\subsection{Hadron Helicity Conservation}

According to pQCD, in any hard-scattering process total hadron helicity should be conserved, i.e.,
$$\Sigma_{initial} \lambda_H = \Sigma_{final} \lambda_H$$
It follows that in the annihilation of nucleon-antinucleon carrying opposite helicities $J=0$ state can not be produced.  Accordingly, while $J/\psi (1^{--}) \leftrightarrow p\bar p$ is allowed, $\eta_c (0^{-+}) \leftrightarrow p\bar p$ is forbidden. Similarly, while \quad $\chi_{1,2}(1^{++},2^{++}) \leftrightarrow p\bar p$ is allowed, \quad $\chi_0(0^{++}) \leftrightarrow p\bar p$ is forbidden.

However, these predictions of forbidden transitions are strongly violated.  We see that $B(\eta_c \rightarrow p\bar p) = 1.2(4) \times 10^{-3}$ is suppressed only by factor 2 compared to $B(J/\psi \rightarrow p\bar p) = 2.12(10) \times 10^{-3}$.  More dramatically, $B(\chi_0 \rightarrow p\bar p) = 41(^{+16}_{-9}) \times 10^{-5}$ is enhanced by a factor 5 compared to $B(\chi_2 \rightarrow p\bar p) = 7.4(10) \times 10^{-5}$.  It is not clear what lies behind these gross violations of the Hadron Helicity Conservation `rule'.

\subsection{QCD at Small $Q^2$}

The QCD sector of the standard model has received impressive experimental support in the perturbative domain of very large $Q^2$. However, the universal truth of QCD, which must include its validity in the small $Q^2$ domain, remains an open question.  Wilczek, for example, has emphasized that, ``if you are interested in quantitative results for $\alpha_s$, (which provide) a quantitative measure of how good pQCD is, there is a large premium for working at small $Q^2$.'' The small $Q^2$, or small mass scale that Wilczek has in mind, is not the scale of $u$- and $d$-quarks, because they present very difficult (e.g., relativistic) problems. It is the scale of $c$-quarks, which are heavy enough, but where $\alpha_s$ is already run to nearly 3 times its value at $m(Z^0)$.

The best way to obtain $\alpha_s(m_c)$ is to make ratios in which the two unknowns, $m_c$ and the wave functions at the origin cancel out. Thus from $B(R_{c\bar c} \rightarrow gg)/B(R_{c\bar c} \rightarrow \gamma\gamma)$ we get $\alpha_s(m_c) = 0.36(7)$ for $\eta_c$, and $\alpha_s(m_c) = 0.36(2)$, for $\chi_2$, for an average value of $\left<\alpha_s(m_c)\right> = 0.36(2)$ 

\begin{figure}
\centerline{\includegraphics[width=3.5in]{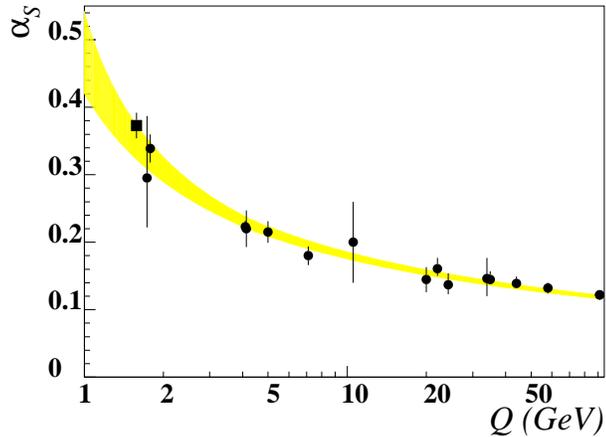}}
\caption{Compilation of world measurements of strong coupling constants (based on ref \cite{pdg}).  The square symbol denotes the result based on charmonium measurements.}
\end{figure}

Figure 4 shows the $\alpha_s$ measurements.  Our $\alpha_s = 0.36\pm0.02$ (at $m_c = 1.5$ GeV) agrees well with $\alpha_s = 0.34\pm0.02$ from $\tau$ decay (at $m_{\tau} = 1.78$ GeV).  Our result corresponds to  $\alpha_s(M_Z) = 0.119 \pm 0.007 \pm 0.007$, whereas the PDG98 average is: $\alpha_s(M_Z) = 0.119 \pm 0.002.$

I must point out that there is a serious caveat associated with the nice value of $\alpha_s$ at $Q_c=1.5$ plotted in Fig. 4.  It has been obtained by using first order radiative correction factors of $\sim2.5$ and 1.9 to the lowest order pQCD predictions for the branching ratios.  Such large factors are admittedly highly suspect.

\section{Future Prospects}

With CLEO III having converted to CLEO-c there is not much prospect of new runs for bottomonium spectroscopy.  Of course, CLEO will continue to mine whatever good bottomonium physics it can from the $\sim5$ fb$^{-1}$ data it has on $\Upsilon(1S,2S,3S)$.  The prospects for charmonium spectroscopy are better, and we can look forward to lots of precision results from CLEO-c running at $\psi'(2^3S_1)$ and $J/\psi(1^3S_1)$.

This research was supported by the U.S. Department of Energy.

Note: The references listed below include several which were not published at the time of the conference, but have become available since then.


\begin{thebibliography}{99}

\bibitem{cleob} CLEO Collaboration, G. Bonvicini et al., \textit{Phys. Rev.} \textbf{D} \textbf{70}(2004)032001.

\bibitem{cleoc} CLEO Collaboration, D. Cronin-Hennessy, \textit{Phys. Rev. Lett.} \textbf{92}(2004)22202.

\bibitem{cleog} CLEO Collaboration, G. S. Adams, et al., \texttt{hep-ex/0409027}, submitted to \textit{Phys. Rev. Lett.}.

\bibitem{pdg} PDG2004, S. Eidelmann et al., \textit{Phys. Lett.} \textbf{B 592}(2004)1.

\bibitem{cleox} CLEO Collaboration, N. E. Adam et al., \texttt{hep-ex/0407028}.

\bibitem{bellea} Belle Collaboration, S. K. Choi et al., \textit{Phys. Rev. Lett.} \textbf{89}(2002)102001; K. Abe et al., \textit{Phys. Rev. Lett.} 89(2002)142001.

\bibitem{cleoa} CLEO Collaboration, D. M. Asner et al., \textit{Phys. Rev. Lett.} \textbf{92}(2004)142001.

\bibitem{babara} BaBar Collaboration, B. Aubert et al., \textit{Phys. Rev. Lett.} \textbf{92}(2004)142002.

\bibitem{elq} E. J. Eichten, K. Lane, and C. Quigg, \textit{Phys. Rev.} \textbf{D} \textbf{69}(2004)094019.

\bibitem{cester} R. Cester, \textit{Proc. Frontier Science 2002}, Frascati Physics Series (2003) 41.

\bibitem{dave} D. Joffe, Ph D. dissertation, Northwestern University, 2004, unpublished.

\bibitem{amiran} A. Tomaradze, Proc. APS meeting of GHP, Fermilab (Oct. 2004), to be published by IOP.

\bibitem{dasp} DASP Collaboration, R. Brandelik et al., \textit{Z. Phys.} \textbf{C1}(1979)233.

\bibitem{besa} BES Collaboration, J. Z. Bai et al., \textit{Phys. Rev. Lett.} \textbf{88}(2002)101802.

\bibitem{cball} C. Edwards et al., \textit{Phys. Rev. Lett.} \textbf{48}(1982)70.

\bibitem{sethb} K. K. Seth, \texttt{hep-ex/0405007}.


\end{thebibliography}
\end{document}